\documentstyle[manuscript,aps]{revtex}
\begin{document}
\draft
\title{A Numerical Study of Phase Transitions Inside the Pores of
Aerogels \footnote{a short report of this work has been published in
Phys. Rev. Lett. {\bf 74}, 422 (1995).}}
\author{Katarina Uzelac}
\address{Institute of Physics of the University,
University of Zagreb, Bijeni\v cka 46, POB 304,
41000 Zagreb, Croatia}
\author{Anwar Hasmy and R\'emi Jullien}
\address{Laboratoire de Science des Mat\'eriaux Vitreux,
Universit\'e Montpellier II, Place Eug\`ene Bataillon,
34095 Montpellier, France}
\date{\today}
\maketitle
\begin{abstract}
Phase transitions inside the pores of an
aerogel are investigated by modelizing
the aerogel structure by diffusion-limited
cluster-cluster aggregation on a
cubic lattice in a finite box and considering
$q$-states Potts variables on the empty sites
interacting via nearest-neighbours. Using
a finite size scaling analysing of
Monte-Carlo numerical results, it is
concluded that for $q=4$ the
transition changes from first order to second
order as the aerogel concentration
(density) increases. Comparison is made
with the case $q=3$ (where the
first order transition is weaker in three
dimensions) and with the case $q=4$ but for randomly
(non correlated) occupied sites. Possible
applications to experiments are discussed.
\end{abstract}


\section{Introduction}

        In recent time a lot of interest has been driven to the study of phase
transitions in porous
materials. Number of experiments have been dedicated to study of new features
of transitions in
helium\cite{1,2,3,4,5}
or liquid crystals\cite{6,7,8,9,10} when they are confined within the
pores of an aerogel. Theoretical reconstruction of those phenomena has been
attempted through various
models\cite{11,12,13,14,15,16}.

This paper presents a numerical simulation in three dimensions for a Potts
model confined within the pores
of a realistic
aerogel structure obtained by simulating a diffusion-limited cluster-cluster
aggregation (DLCA) process.
In spite of its simplicity, the Potts model hides a number of differently
behaving models (such as the
Ising model, percolation, ...) which can be obtained by a particular choice of
the number of states $q$.
In three dimensions it exhibits a first order transition when $q \geq 3$.

The aim of the present article will be to examine how this transition is
modified when the studied
fluid is confined within pores of an aerogel. Previous studies of the two
dimensional Potts model
in a randomly diluted medium \cite{17,18,19} show crossover from first order to
second order type behaviour.
The same effect was also discussed in more a general context by Berker
\cite{20}.

Our study will focus on the 4-states Potts model, for which the first order
character of the transition
is strongly pronounced.
Analogous  calculations will be performed in two other cases for comparison:
$q=3$ within an aerogel and $q=4$ within randomly diluted medium.


\section{Constraints on theory}

        The Potts model and the aerogel  are described here on a $LxLxL$ cubic
lattice.
The aerogel structure has been obtained by using the diffusion limited
cluster-cluster aggregation (DCLA)
process  with periodic boundary conditions (PBC)\cite{21}. This model is
precisely the lattice version of the
off-lattice model extensively presented elsewhere\cite{22}. The same lattice
version has been recently extended
to investigate numerically the sintering of aerogels\cite{22}.
The fraction of sites occupied by the aerogel has been varied from $c=0.05$ to
higher concentrations.
In all the configurations used here we have taken care that the aerogel
structure is connected in all directions.

         On all the remaining free sites of the lattice are positioned
$q$-states Potts variables
$\sigma_i$ which are interacting via nearest-neighbour interactions given by
the Hamiltonian:
\begin{eqnarray}
H = -J\sum_{<ij>}\delta_{\sigma_i \sigma_j}
\label{E1}
\end{eqnarray}
where $\delta$ is the Kronecker symbol. The summation is taken over all the
free sites and includes PBC.
It has been assumed here that the Potts variables do not interact with the
aerogel. Consequently the effect
of aerogel is understood as a particular kind of specially correlated dilution
with concentration $c$.
This gives the motivation to perform also the calculations in a randomly
diluted medium of equal concentration
in order to separate the effects of pure disorder from those due to the
particular correlations existing in
aerogel.

        In the present work we have performed Monte-Carlo (MC) simulations by
using a simple spin flip
Metropolis algorithm. Box sizes of $L$=10,12,15 and 20 have been considered.

        The basic quantity that we study and analyse here is the energy
probability distribution defined as
\begin{eqnarray}
P_L(E) = {1\over Z_L(K)}N_L(E)\exp{(-KE)}
\label{E2}
\end{eqnarray}
where $K=J/T$, $E$ is the energy in units of $J$,
$Z_L(K)$ is the partition function and $N_L(E)$ is the total number of
configurations with energy $E$
for the system of dimension $L$. The shape of  $P_L(E)$ is different for
first or second order phase transitions. A first order transition, in the
vicinity of $T_c$, is characterized by a double peak structure in
$P_L(E)$ resulting from the coexistence of the ordered and disordered phases.

The depth of those maxima is related to the interface free energy $\Delta F$
between the ordered and
disordered phases:
\begin{eqnarray}
\Delta F_L = \ln{P_L(E_M)\over P_L(E_m)}
\label{E3}
\end{eqnarray}
where $E_M$ and  $E_m$ denote the energy of one of the maxima and the energy of
the minimum of $P_L(E)$.
respectively.
For a second order transition, on the contrary, $P_L(E)$ exhibits only one
maximum.
Recently it was shown \cite{24} that a finite size scaling (FSS) analysis of
$\Delta F$ can be a very
efficient tool to detect a first order phase transition, even in the cases
where the scaling is performed
for sizes lower than the the finite correlation length of the first order
transition problem. Besides
the Binder's fourth order cummulant\cite{25} the study of $\Delta F$ scaling
has since then been successfully
applied to number of problems \cite{26} as a criterion for a first order
transition. When the transition is
of the first order, the interface energy $\Delta F_L$ should scale as
$L^{d-1}$, while otherwise it is
expected to disappear
with increasing $L$.
The gradual disappearance of the two maxima with increasing concentration and
their scaling in $L$ are illustrated on fig.(1) for the case of 4-state Potts
model within aerogel of
different concentrations. The analysis of $\Delta F$ will be given later.

The maxima of $P_L(E)$ are also used to determine the critical temperature. We
define by $T_c(c,L)$
the point where the two minima are of equal height. $T_c(c)$ is then obtained
by making the extrapolation
to infinte $L$.

Simulations were performed to obtain the best possible guess for $T_c$ (usually
closer than 0.01\%). Then
for close inverse temperatures $K'$, $P_L(E)$ were found by using the histogram
method of Ferenberg and Swendsen
\cite{26} based on the formula
 \begin{equation}
P_{L}(E,K')=\frac{P_L(E,K)e^{(K'-K)E}}{\sum_EP_L(E,K)e^{(K'-K)E}}
\label{Efs}
\end{equation}
which enabled a more precise determination of $T_c$, without additional MC
runs. The histogram method is very
sensitive to the degree of ergodicity achieved.
Our simulations have been done with $10^6$-$4*10^6$ Monte Carlo flips per spin
(MCS). The ergodicity of the
results has been examined by calculating also the probability distribution for
the different components of
the "magnetisation" $P(\alpha,m)$, where $m$ is defined as
\begin{equation}
m(\alpha)=\frac{qM(\alpha)-1}{q-1}  ~~~~~~ \alpha=1,...,q
\end{equation}
and where $M(\alpha)=<\sigma(\alpha)>$ is the average of the $\alpha$-component
of the order parameter $\sigma$.
By comparing $P(\alpha,m)$ for different components one can check if the system
has visited all parts of the
phase space.
On fig.(2) we present a typical example obtained after $2*10^6$ MCS where the
graphs for different components
are hardly distinguishable.

More information can be obtained for $P_L(E)$ by calculating higher moments of
the energy distribution. The moment of order $n$ is defined as:
\begin{eqnarray}
<E^n>_L = \sum_E E^n P_L(E)
\label{E6}
\end{eqnarray}
Here we shall present the results for the specific heat, which has been
calculated from
\begin{eqnarray}
C_L(T) = {L^3\over T^2}(<E^2>_L - <E>_L^2)
\label{E5}
\end{eqnarray}
The specific heat $C_v$ also exhibits drastically different behaviours in the
two cases.
For a first order transition $C_v$ behaves as a $\delta$ function so that it
scales as $L^d$. For a
second order transition it scales as $L^{\alpha/\nu}$, where the $\alpha$ and
$\nu$ are the specific heat
and correlation length critical exponents respectively.

Our study is centered around the $q=4$ case of the Potts model within the pores
of an aerogel.
We have chosen $q=4$ since this model exhibits in the pure case a strong first
order transition.
For comparison we have also studied two other cases.
First is the $q=3$ case within an aerogel. For $q=3$ the first order character
of the transition
is very weak. For illustration, we have reported in fig. (3) $P_L(E)$ at $T_c$
for $q=3$ and $q=4$.
The aim was to to examine how the results would change with changing the number
of degrees of freedom of
the considered model.
Second is the $q=4$ case with random dilution in the same range of
concentrations already discussed. When comparing
with the aerogel case, we hope to see the influence of the special short range
correlations that exist in
aerogels.
Before presenting the results, let us stress that the present study is based on
simulations made for
only a few configurations per given concentration, and no configurational
average has been performed.
For this reason, the results obtained by the scaling analysis are not expected
to give precise quantitative data,
especially as it concerns the critical exponents.

\section{Results}

\subsection{Phase diagram  (Critical temperature)}

        Three sets of results for $T_c(c,L)$ calculated from $P_L(E)$, (as
explained in the previous section)
are presented in fig. (4). They cover the cases $q=4$ in aerogel, $q=4$ with
random dilution and $q=3$ in aerogel.

There we show the data for $L=15$ and $L=20$. In all the three cases the data
for $T_c$ depend rather weakly on $L$.
Their variations with $L$ are comparable to the error bars which are estimated
to be of order $10^{-4}$ for
$c < 0.2$ and $10^{-3}$ for the higher concentrations presented here. When
considering the lower concentrations,
the error bars are determined in first place by the precision of the MC
procedure and the histogram method which
depends on the degree of ergodicity observed.
For higher concentrations the configurational fluctuations of disorder are
expected to bring the main contribution
to the error bars.

         Comparing $q=4$ with $q=3$ results, both in aerogel, we find that the
normalisation of $T_c$(c) by
$T_c$(0) completely eliminates the difference between the two cases (the
dilution has the same effect on our model,
independently of the number of degrees of freedom $q$).

Comparison of the $q=4$ model in aerogel with the $q=4$ case with random
dilution shows, in contrary, a very clear
difference. The  fact that $T_c$ for the case with
random dilution is lower than $T_c$ in aerogel can be easily understood from
geometrical arguments.
In the random dilution case a larger number of bonds between Potts spins has
been suppressed than in the aerogel
case, where all the "dilution" sites are connected.
If we neglect the fluctuations, we can try to relate $T_c$ to the effective
coordination number $<z>$ between
Potts sites using the expression
\begin{eqnarray}
{T_c(c)\over T_c(0)} = {<z>\over 6}
\label{E4}
\end{eqnarray}
For the random dilution case, $<z>=6(1-c)$, which gives a linear dependence on
concentration $c$, illustrated on
fig. (4)  by the plain line. For aerogel, it can be calculated numerically and
gives a nonlinear dependence
(presented on the figure by the dashed line).
In spite of neglecting the fluctuations, the agreement of numerical data with
eq.(\ref{E4}) is surprisingly good.
The largest discrepancy is obtained in aerogel for points between $c=0.1$ and
0.2 where one expects to find a
crossover from  first to second order regime.

\subsection{Interface energy}

The results for the scaling of the interface free energy are summarized on fig.
(5) for the three cases considered
above.
The precision of the points is  at least one order poorer than for $T_c$.

        As it could be concluded from figs. (1a-c), $\Delta F$ disappears for
$c\geq 0.2$ so that here we present
the behaviour for three different concentrations: $c=0.05$, 0.1 and 0.15
compared to the pure case.
The plot of $\Delta F_L$ versus $L^2$  should be a straight line if the
transition is of first order.
Let us consider first the aerogel $q=4$ case. As expected, a straight line is
obtained for the pure case.
A similar behaviour is obtained for  $c=0.05$. For $c=0.1$
and $c=0.15$ the results appear to be strongly configuration dependent,
fluctuations of disorder becoming important
close to the crossover.
More intensive calculations implying a larger number of configurations would be
needed to determine the exact
position of the crossover concentration.

        The 4-states case with random dilution shows, even qualitatively, a
very similar behaviour than
the aerogel case.




For $q=3$, as already seen on figs. (3), the interface energy is lower, and the
effect of disorder seems to manifest
itself at lower concentrations. Thus, we can expect the crossover concentration
to be dependent of $q$ more than on
the type of disorder.

\subsection{Specific heat}

        The specific heat exhibits a well pronounced maximum, which becomes
more broadened as the aerogel
concentration increases (fig. (6)). On fig. (7) we try to examine the scaling
of $C_{Lmax}$ in the $q=4$ case
in aerogel, on a log-log plot, limited by the fact that, in the case of the
specific heat, the data are even
less precise than for $\Delta F$, so that at this stage we expect to obtain
only a qualitative idea when
comparing various concentrations. The pure case gives the most clear results
with a slope very close to 3
as expected for a first order transition. For other concentrations the
fluctuations become much more important.
One can observe however for $c=0.05$ a slope still close to 3, while for higher
concentrations it becomes much
smaller as an indication for a different regime.

\section{Discussion}

        In the preceding sections we have presented results of Monte Carlo
simulations together with a
scaling analysis for the Potts model on a cubic lattice either occupied by an
aerogel or randomly diluted
with the same concentration rates.

In spite of the limitations imposed by the present level of calculations, where
no averaging over disorder
configurations has been done, we can rise a few points of discussion.

        The critical temperature $T_c(c)$ can be rather well approximated by
normalizing the pure system
critical temperature $T_c(0)$ by the effective coordination number for both
cases of disorder. Consequently,
there is a clear difference between the phase diagram curves $T_c(c)$ for
aerogel and random dilution.
        It has also been found that there is a nonzero crossover concentration
at which the first order transition
changes to a second order one, induced by disorder. This change does not seem
to be much influenced by the
details of dilution (such as correlations), but depends on the number of Potts
states $q$.
Question is if the disorder type would remain of so little relevance if we
would take into account
a more realistic situation where the Potts spins interact with the aerogel.

More intensive calculations, involving  averaging over different
configurations,
would be needed to determine with precision the crossover concentration and
calculate the critical exponents
in the second order regime.
The calculations for the Potts model with random dilution that have previously
been performed in
two dimensons \cite{17,18,19}
show that critical behaviour in the second order regime is the same as that of
the Ising model. Our preliminary
results for the specific heat support the thesis that this might be also true
in the aerogel case in three dimensions, but more
funded calculations are needed.

\section{Conclusion}

To conclude, we would like to stress that the results of the present
calculation are in good qualitative
agreement with the general features of most of the existing experiments. In the
case of $^4$He, it has been
observed that the decrease of $T_c$ (compared to the pure case) is more and
more pronouced when it is
diluted in aerogel, xerogel, and vycor, successively \cite{1}. This is in
general agreement with our phase
diagram which predicts a monotonic decrease of $T_c$ with increasing aerogel
density. Moreover the observed rounded
specific heat and the detected second order nature of the phase transition of
nCB
liquid crystals in a relatively dense aerogel (with density corresponding to
$c=0.16$)\cite{7}
are in agreement with our general conclusion
for a change from first order to second
order above a given crossover aerogel density.

We would like to suggest a systematic
series of experiments of phase transitions in the pores of gradually densified
aerogels in order to get an
explicit experimental variation of the critical temperature with the aerogel
density and to detect the
existence of the density crossover that we predict here.

Of course the use of the Potts model limits the
possibilities of quantitative applications of our results. In the future we
intend to use different
hamiltonians more specifically adapted to the existing experimental situations.

\begin{figure}
\caption{Energy probability distribution at critical temperature $T\_c(c,L)$
for $q=4$
 in aerogel boxes of size $L=12$ ( plain line ), $L=15$ (dotted line),
$L=20$ (dashed line) for three different concentrations: $c=0.05$ (a), $c=0.15$
(b) and
 $c=0.3$ (c).}
\end{figure}
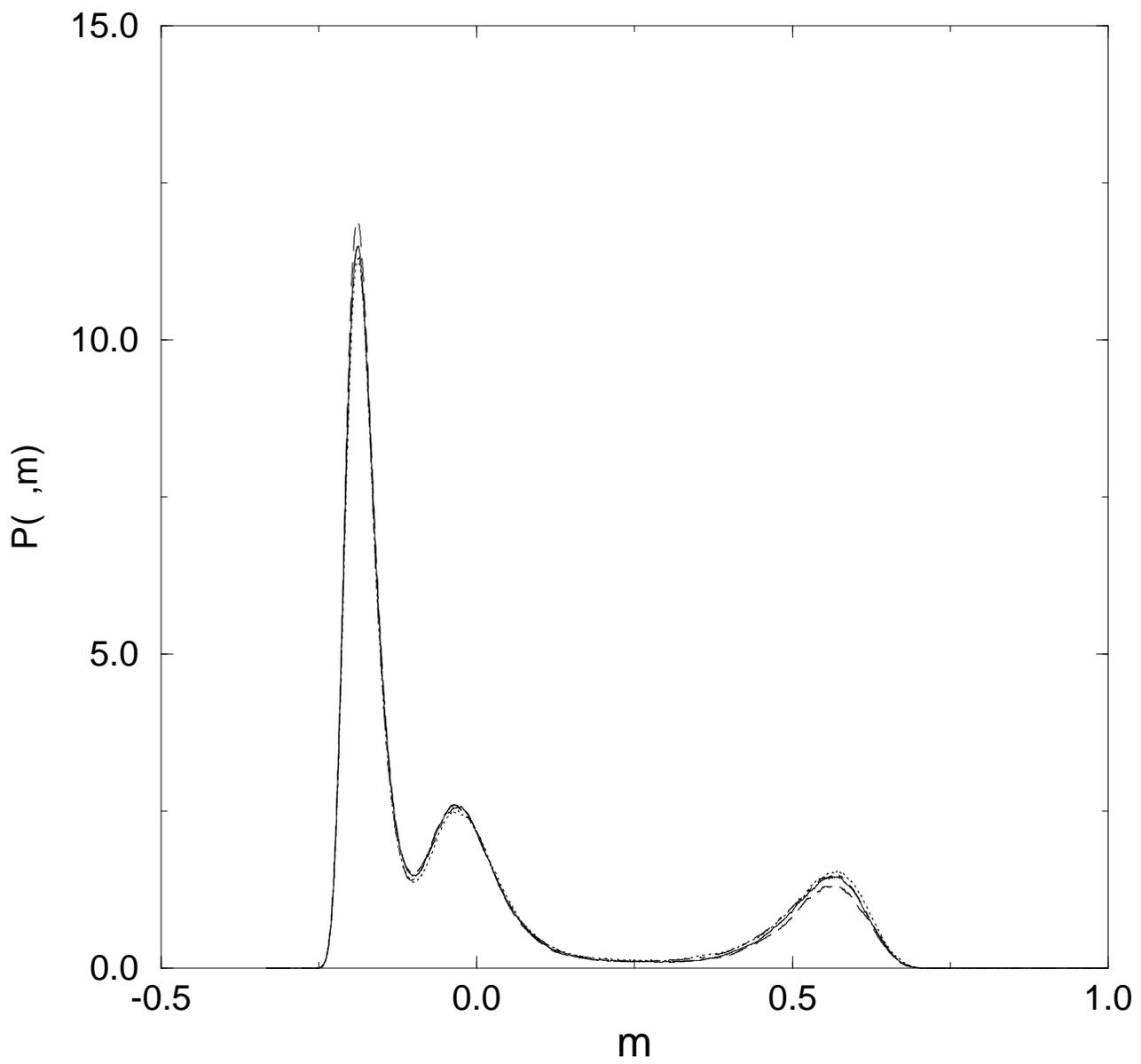
\begin{figure}
\caption{Probability distribution for the four components of the
order parameter in the aerogel case with $q=4$, $c=0.05$, $L=15$, $T=1.5568$
after $2*10^6$ MCS.}
\end{figure}
\begin{figure}
\caption {Energy probability distribution at $T_c$ in the pure case ($c=0$) for
sizes
$L=12$ (plain line), $L=15$ (dotted line), $L=20$ (dashed line)
for $q=3$ (a) and $q=4$ (b).}
\end{figure}
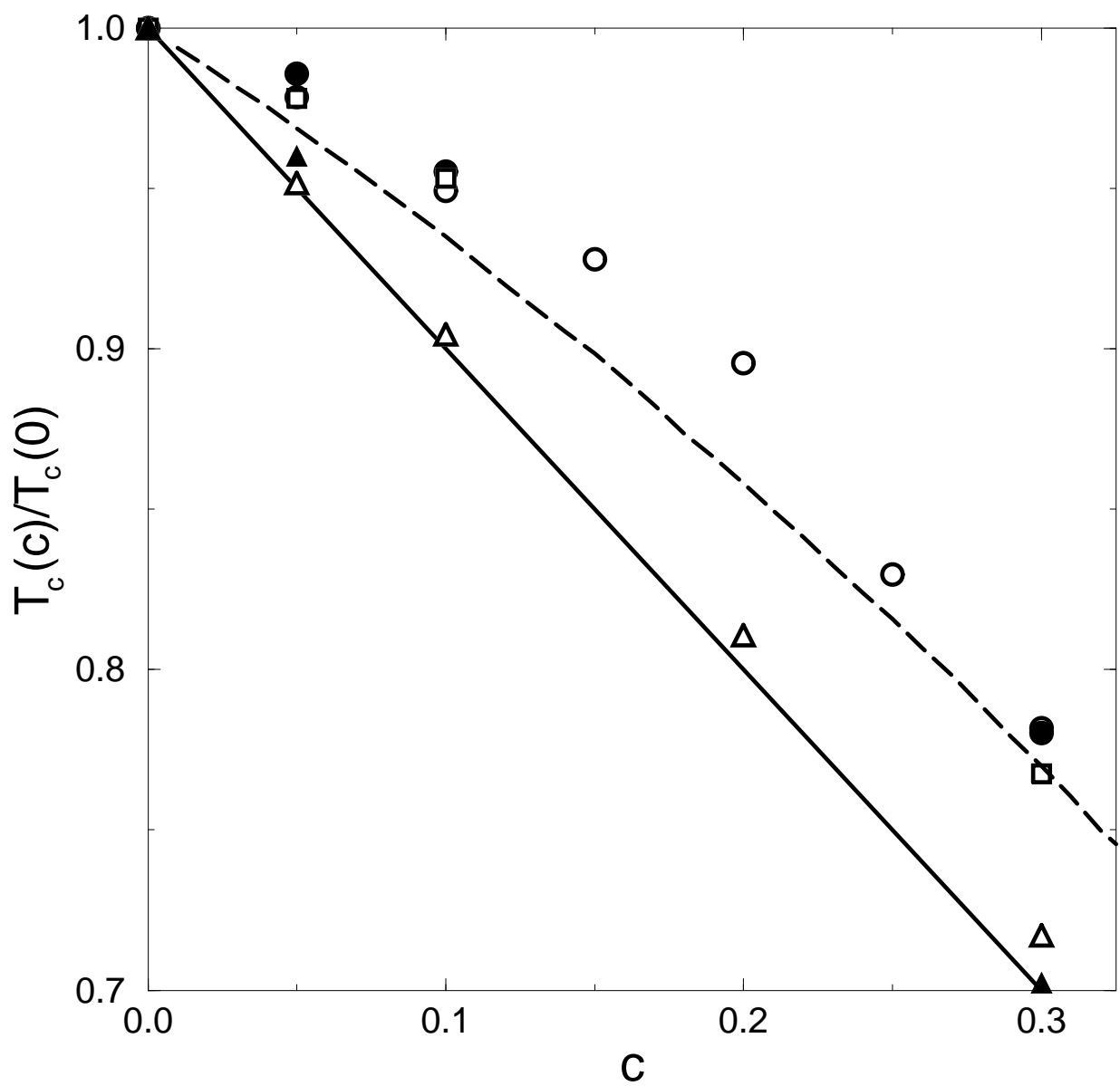
\begin{figure}
\caption{The reduced critical temperature $T_{c}(c)/T_{c}(0)$ is plotted as
a  function of $c$ for three cases: $q=4$  within aerogel (circles), $q=3$
within aerogel
(squares) and $q=4$ with random dilution (triangles). Open symbols correspond
to
$L=15$ and dark symbols to $L=20$ results. The dashed and solid lines
correspond to
the approximation using eq.(8) for the aerogel and the random dilution case
respe
ctively.}
\end{figure}
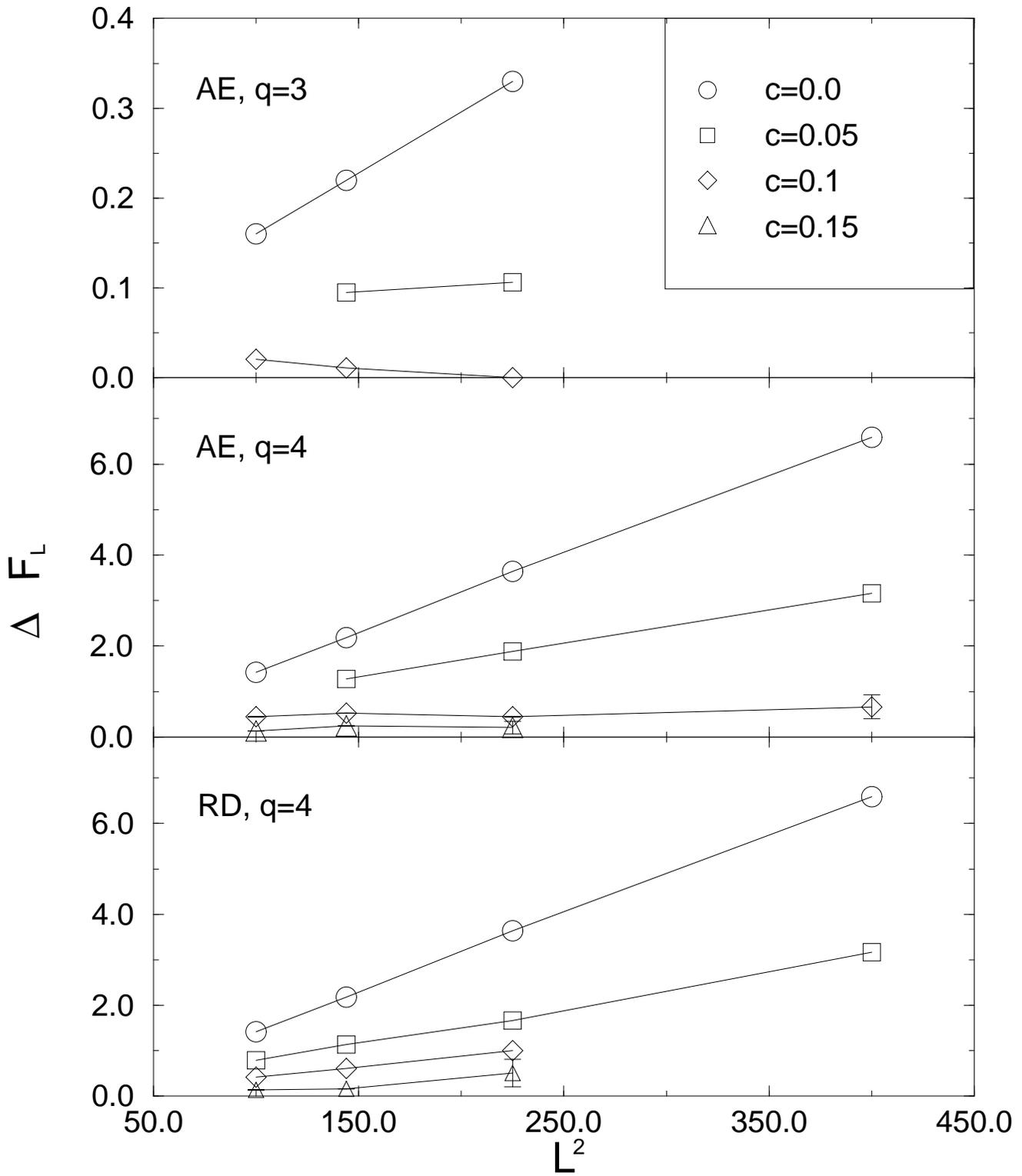
\begin{figure}
\caption{The interface free energy $\Delta F_L$ is plotted as a function of
$L^2$ for different $c$-values.
Top, middle and bottom correspond to ($q=3$, aerogel), ($q=4$,
aerogel), ($q=4$, random dilution), respectively.}
\end{figure}
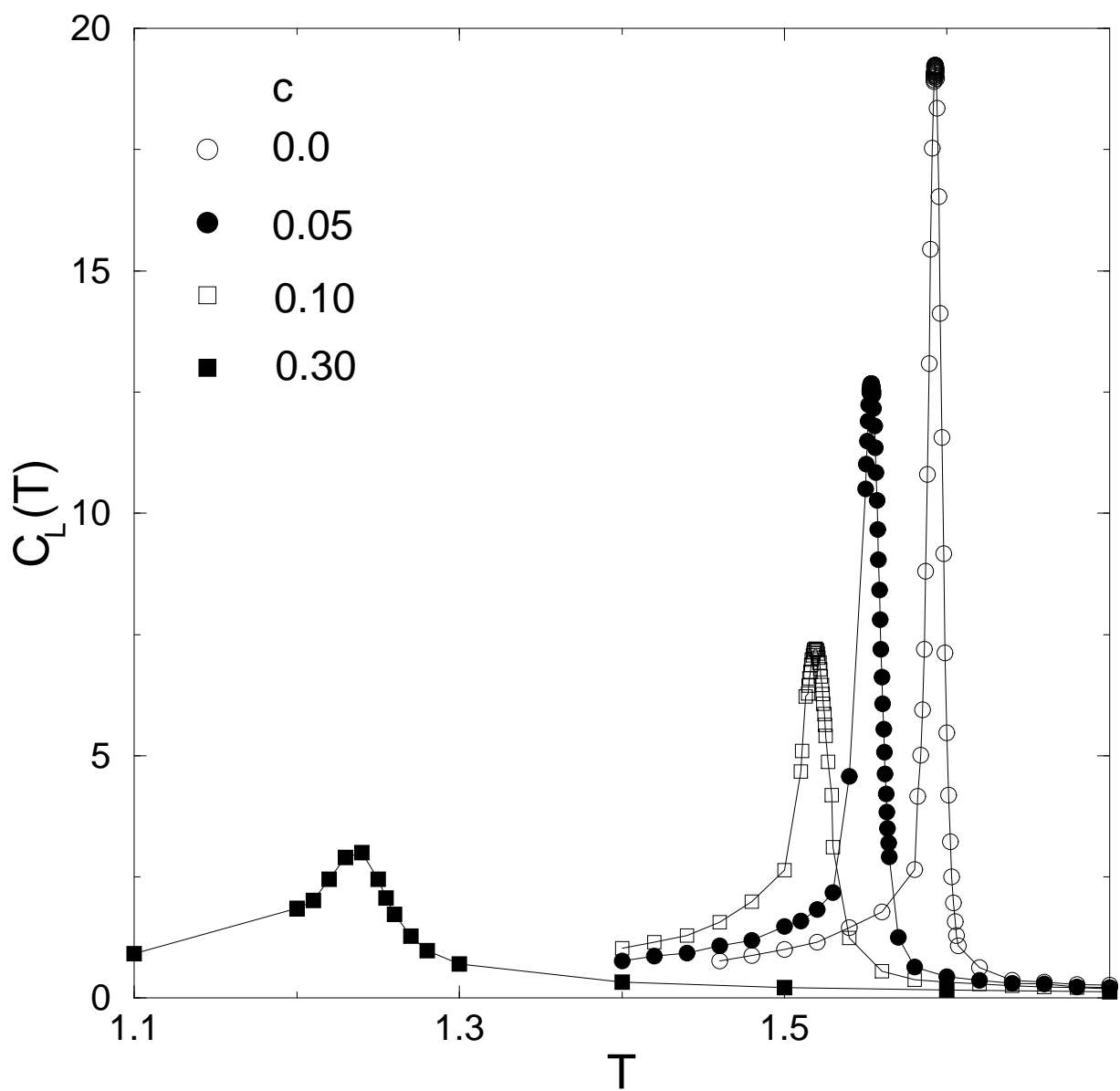
\begin{figure}
\caption{The specific heat $C_L(T)$ is plotted as a function of $T$ in
the aerogel case for $q=4$, $L$=12 and for different $c$ values.}
\end{figure}
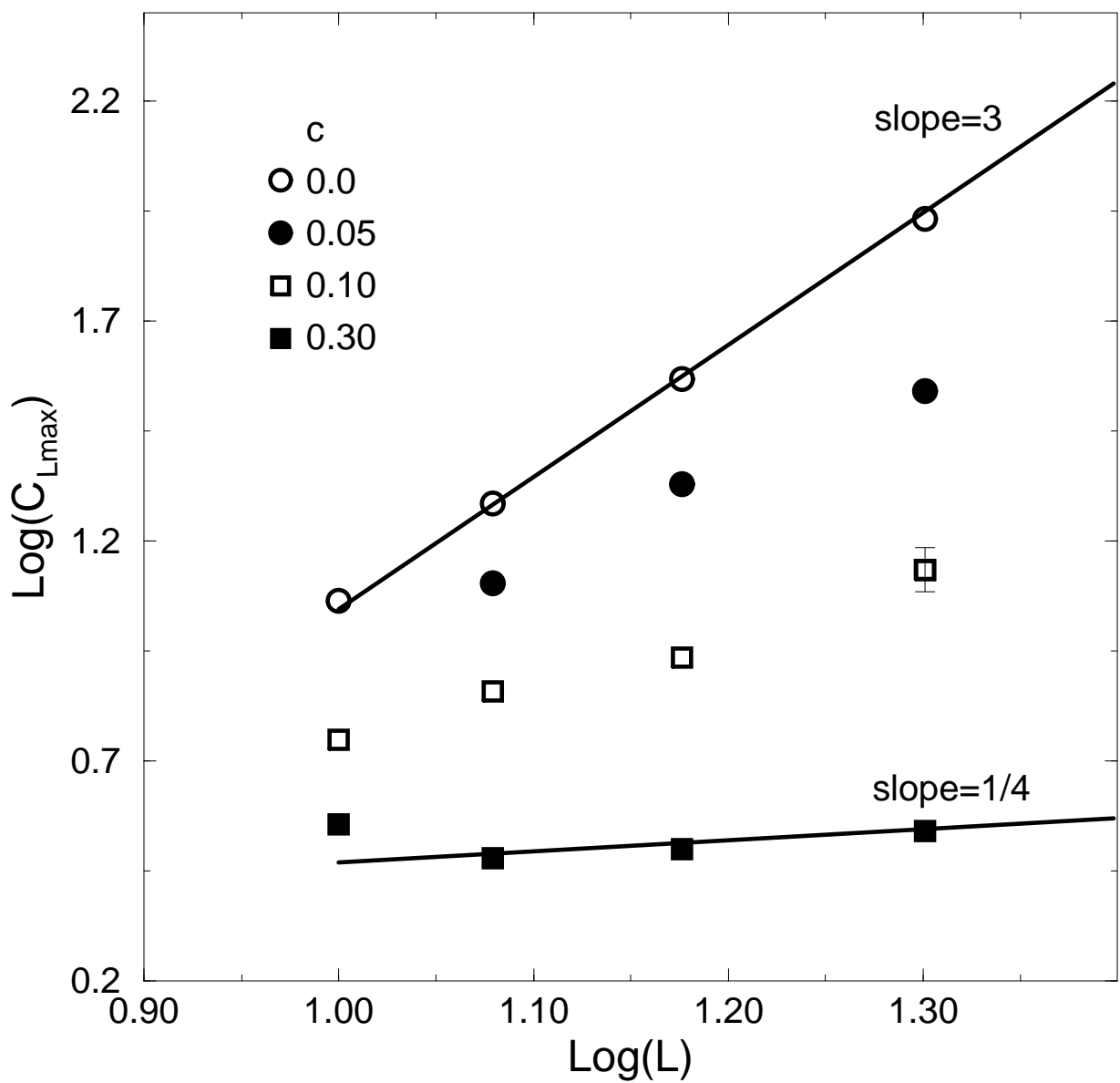
\begin{figure}
\caption{Log-log plot of $C_{Lmax}$ versus $L$ in the aerogel case
for $q=4$ and for different $c$ values.}
\end{figure}

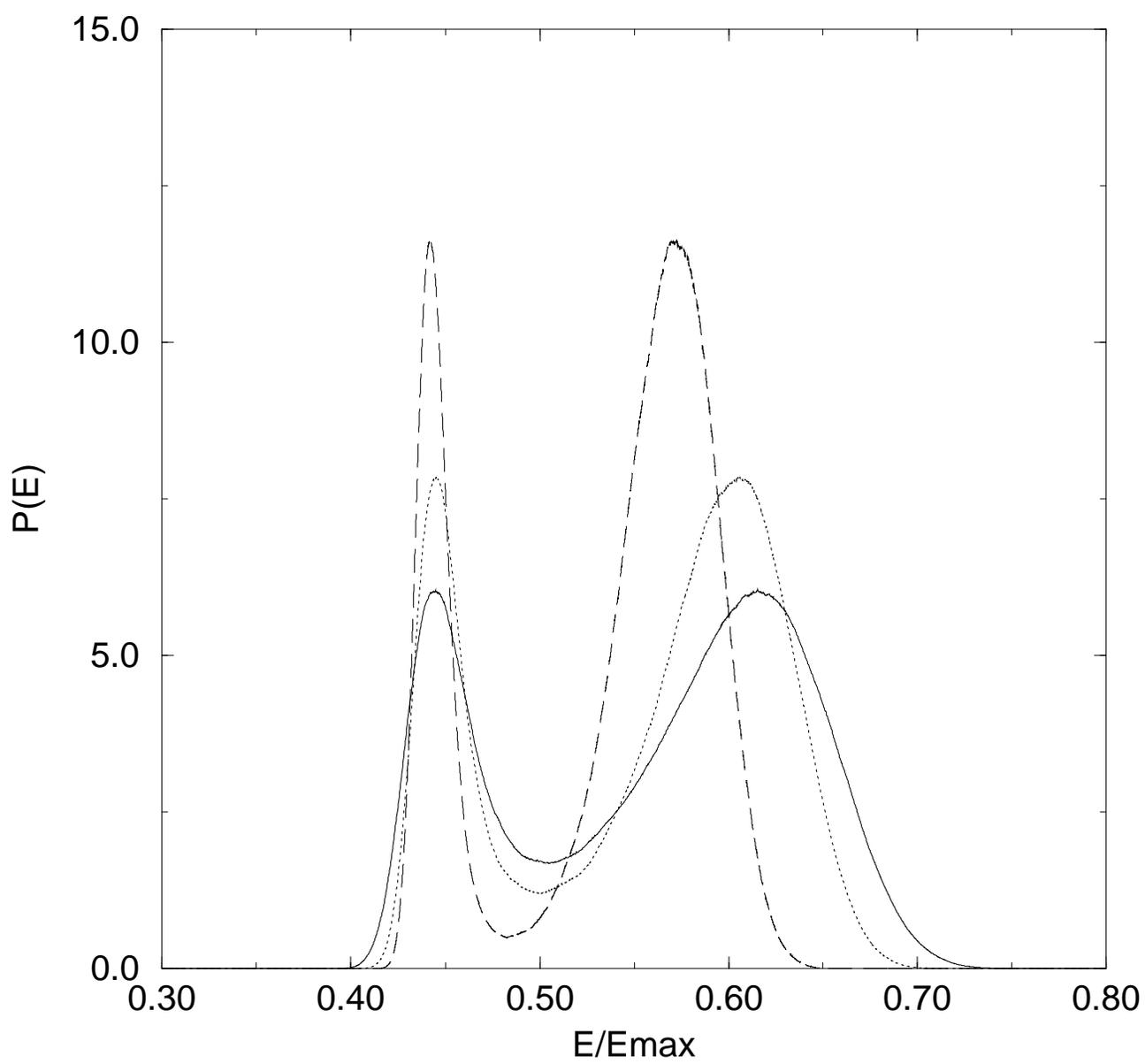
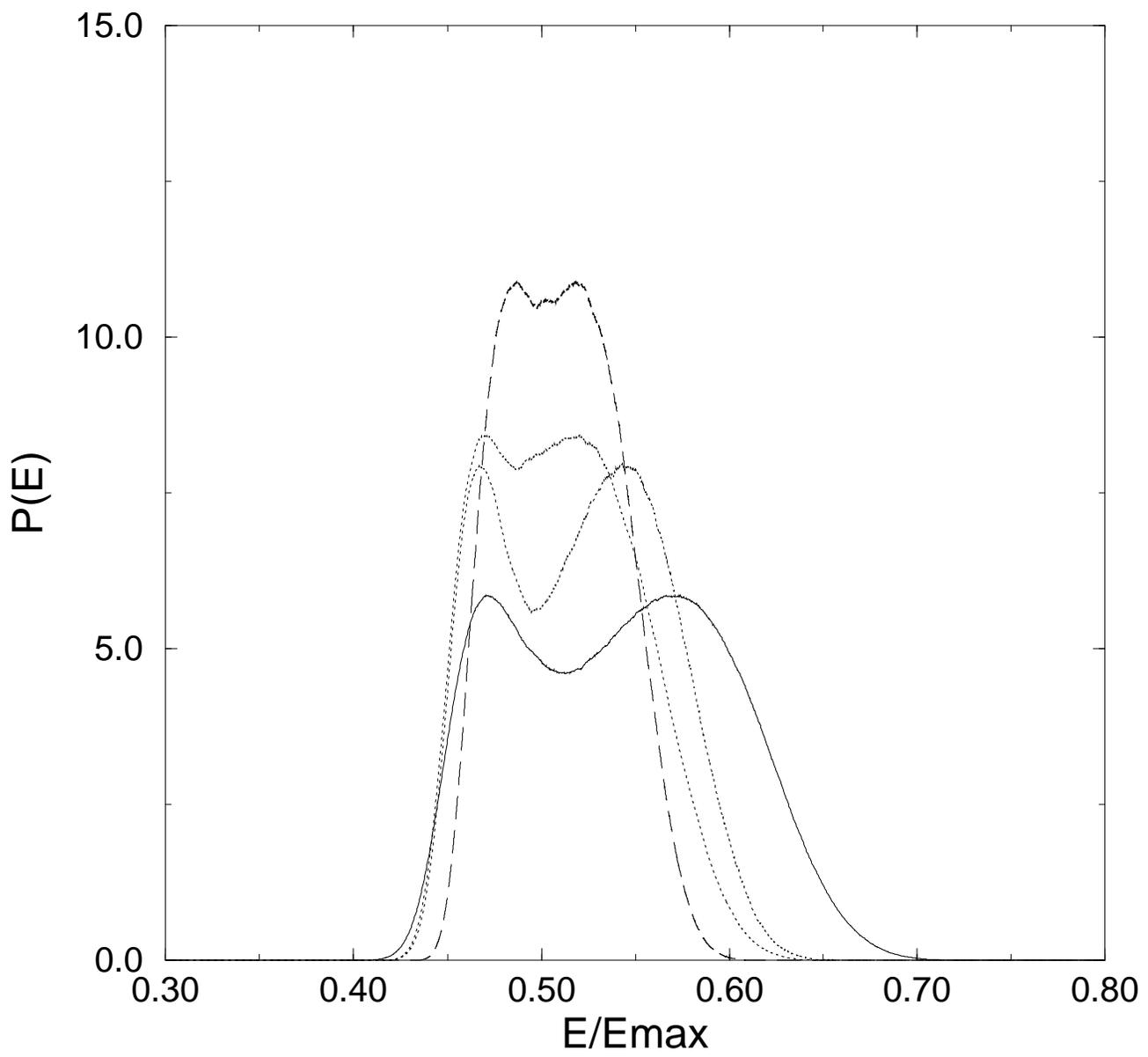
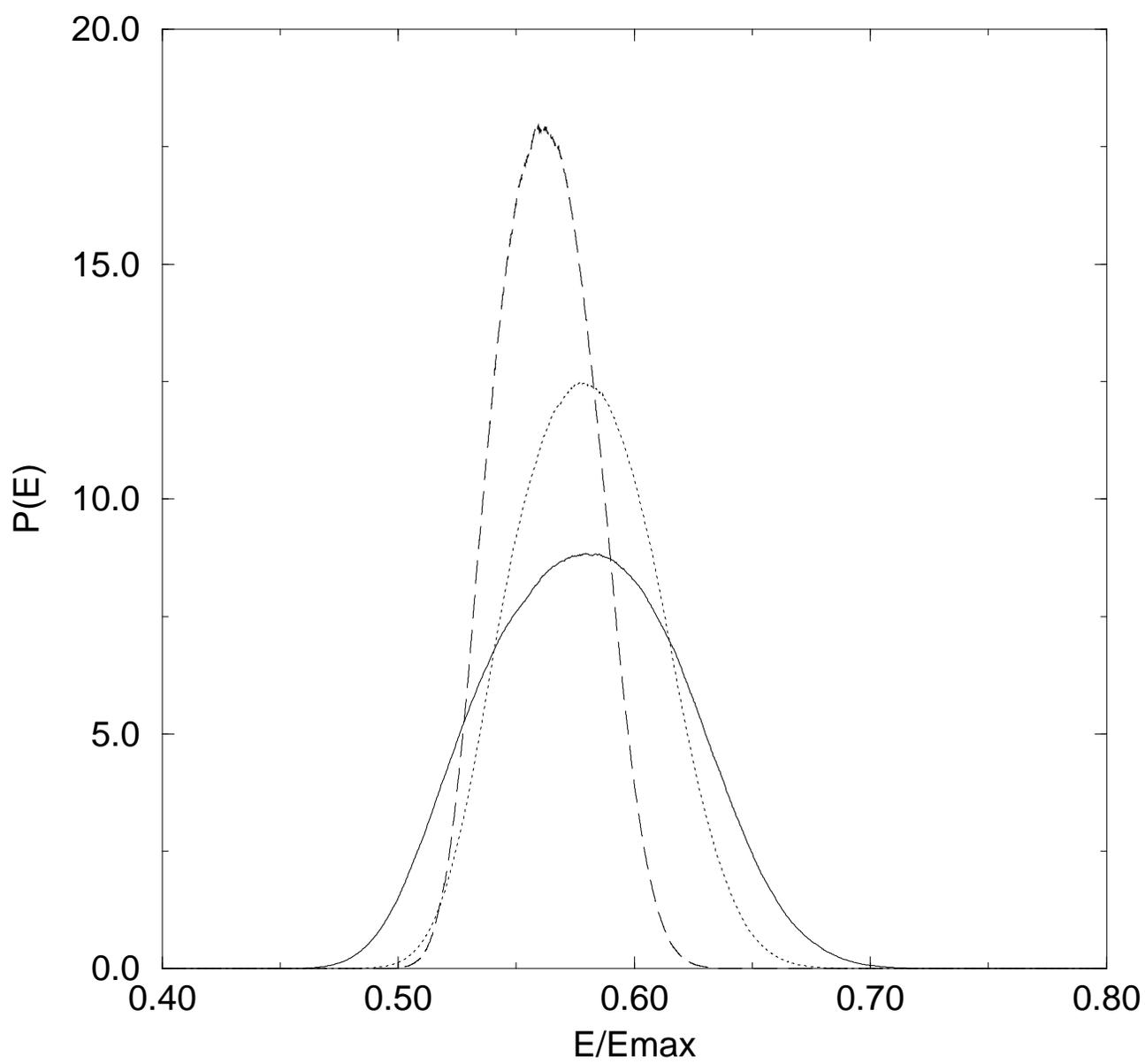
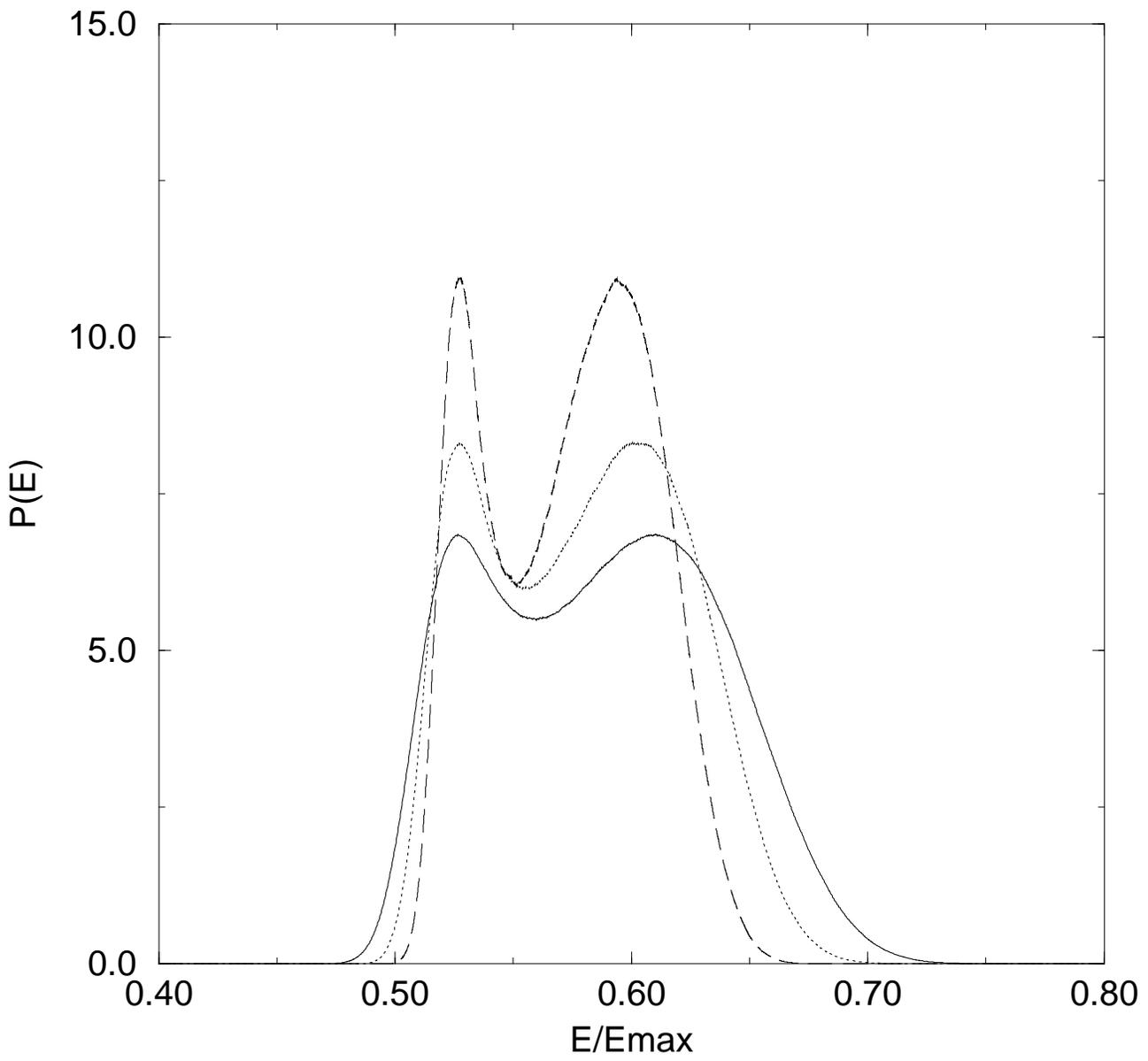
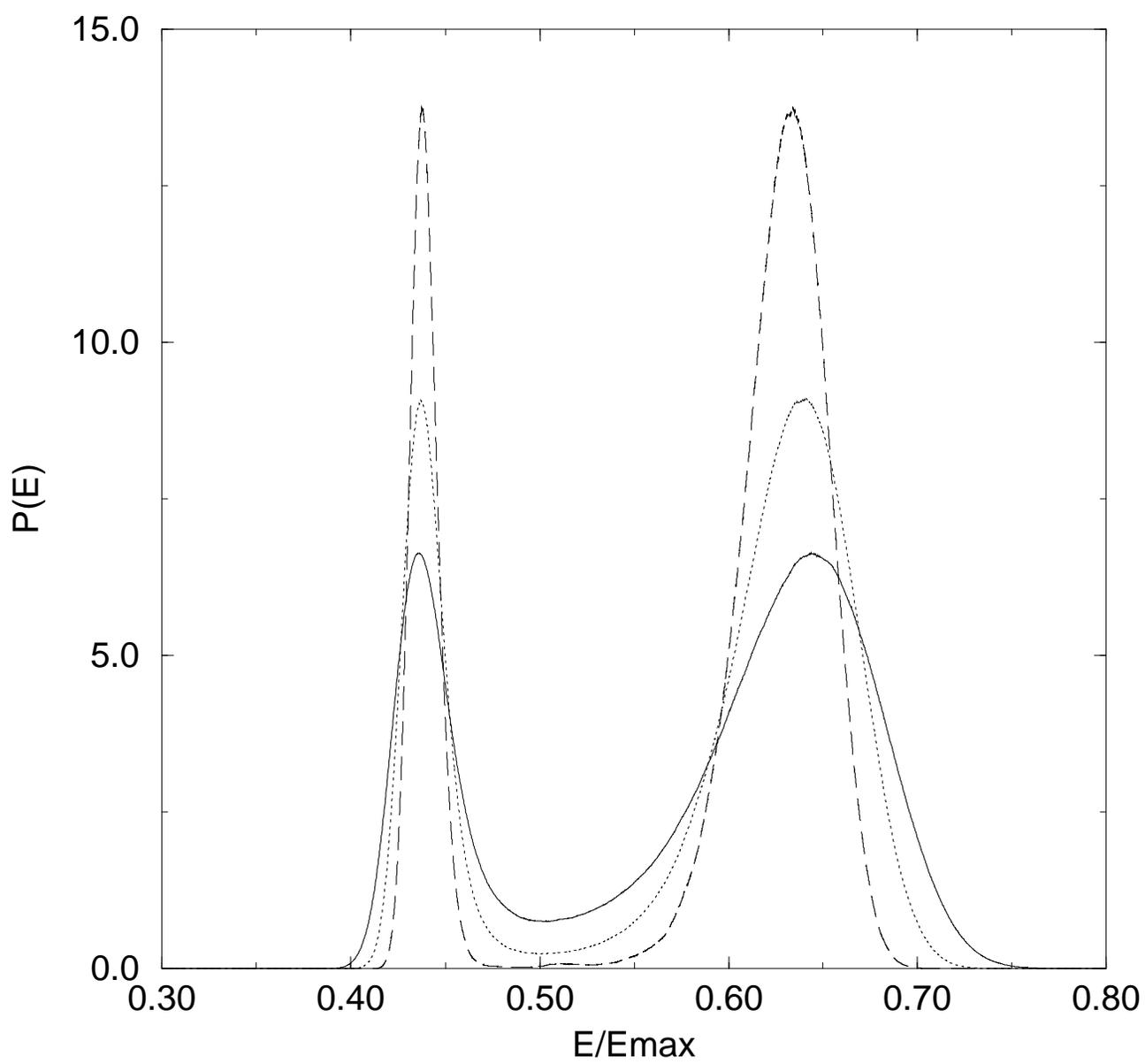
\end{document}